\documentstyle[12pt]{article}
\normalsize

\begin{document}
\title{A New Class of Bounds for Correlation Functions \\
in Euclidean Lattice Field Theory and Statistical Mechanics of Spin
Systems}
\vspace{4cm}
\author{Manfred Requardt \\ Institute for Theoretical Physics \\
University of G\"ottingen\\ Bunsenstra{\ss}e 9\\ D-37073 G\"ottingen
\\ Federal Republic of Germany}
\vspace{3cm}
\date{July 1994}
\vspace{3cm}
\maketitle
\vspace{3cm}
\begin{abstract}
Starting from an extension of the Poisson bracket structure and
Kubo-Martin-Schwinger-property of classical statistical mechanics of
continuous systems to spin systems, defined on a lattice, we derive a
series of, as we think, new and interesting bounds on correlation
functions for general lattice systems. Our method is expected to
yield also useful results in Euclidean Field Theory. Furthermore the
approach is applicable in situations where other techniques fail,
e.g. in the study of phase transitions without breaking of a {\bf
continuous} symmetry like $P(\phi)$-theories with $\phi (x)$ scalar.
\end{abstract}

\newpage
\section*{1. Introduction}
\setcounter{equation}{0}
\setcounter{section}{1}
In more recent times it has become more and more apparent that the
natural Poisson bracket structure of systems of classical point
particles is of considerable conceptual {\bf and} practical value in
the investigation of all sorts of questions being related to e.g.
phase transitions in thermal equilibrium systems, transport equations
and the like (see e.g. refs. [1] to [8]).

One reason of its usefulness derives from the fact that both the
classical equilibrium condition (more specifically, the so-called
Kubo-Martin-Schwinger (KMS)-condition; cf. e.g. [9]) {\bf and} the
concept of symmetry (breaking) can be very neatly implemented by this
structure.

The KMS-condition in the regime of classical statistical mechanics of
point particles reads:
\begin{equation}
<\{A,B\}> = \beta \cdot <B \cdot \{A,H\}>
\end{equation}
with $A,B$ local observables (without loss of generality: real,
differentiable functions with compact spatial support on phase
space), $H$ the Hamiltonian, $\beta$ inverse temperature, $<\cdot>$
the thermal average and the Poisson bracket given by
\begin{equation}
\{A,B\} = \sum_i (\partial_{r_i} A \cdot \partial_{p_i}B -
\partial_{p_i} A \cdot \partial_{r_i}B)
\end{equation}
(1.1) was already derived by Mermin (cf. ref. [10]) and was used by
him
to get the classical counterpart of the so-called Bogoliubov
inequality
\begin{equation}
<\{A,B\}>^2 \; \le \beta <A^2><\{B,\{B,H\}\}>
\end{equation}
which underlies, in various disguises, some of the approaches
developed e.g. in the above cited papers.

On the other side there exists a considerable amount of model systems
in classical statistical mechanics which do not openly carry such a
nice structure, to mention typical cases in point, systems living on a
discrete base-manifold as e.g. spin systems or systems defined via
functional
integrals.

In view of the great calculational advantages of the Poisson bracket
formalism in classical statistical mechanics it would be highly
desirable to have an analogous machinery at ones disposal for such
(non-canonical) systems.

The necessary general steps in this direction have been undertaken by
us in ref. [11]. Furthermore we studied a (however rather limited)
class of models (e.g. interface Hamiltonian) to demonstrate the
usefulness of our approach but made no systematic analysis of more
complicated classes of spin systems which, while being of genuine
relevance in statistical mechanics as such, serve furthermore as the
starting point of Euclidean quantum field theory.

That is, we will derive in the following in a more systematic manner
various (possibly new) classes of correlation (in)equalities for
spin systems and apply them to typical problems in this field.

\section*{2. The Basic Estimates}
\setcounter{equation}{0}
\setcounter{section}{2}
In this chapter we will apply our general results ([11]) to spin
systems. Furthermore, to keep matters transparent, we will treat, in
a first step, continuous spin systems, i.e. each spin $S(x), \; x \in
{\bf Z}^d$, ranges from plus to minus infinity.

This is a sufficiently large class of models being frequently
discussed in the literature. Our formalism can of course also be
applied to spins living on, say, a compact manifold as e.g. $S^n$
etc.

In that particular case the independent variables defining the
configuration space would be certain angles. With $\{S(x)\} = {\bf
R}$ the
technical manipulations are a little bit simpler (for a discussion of
e.g. the $x-y-$model along these lines cf. [11]).

\noindent{\bf 2.1 Remarks:} i) One can, at least in principle, get
corresponding results for constrained spin systems as limiting cases
by approximating their support, e.g. given by $\delta (S^2-1)\cdot d
S, S \in {\bf R}^d$, with the help of a sequence of smeared out
distributions, taken from som function space, $\rho_n (S^2 - 1)$,
which converge in this limit towards $\delta(S^2-1)$.

ii) Technically it turns out to be advantageous to absorb a possible
extra weight function $\rho(S)$ occurring in the single-spin
distribution into the Hamiltonian by writing it as an exponential.

With the base space being ${\bf Z}^d$ or $a \cdot {\bf Z}^d, a$, the
lattice
spacing, statistical mechanics on a subset $\Lambda \subset {\bf
Z}^d$ is
defined in the usual way via a Hamiltonian $H (\underline S),
\underline S
:= \{S (x_i)\}, x_i \in \Lambda$, i.e.:
\begin{eqnarray}
<A> &:=& Z^{-1} \cdot \int A(\underline S) e^{-\beta H(\underline S)}
d \underline S \nonumber\\
\mbox{with } d \underline S &:=& \Pi_{\textstyle x_{i} \in \Lambda} d
S
(x_i),\nonumber\\
Z &:=& \int e^{-\beta H (\underline S)} d \underline S
\end{eqnarray}
(possible extra weight factors being absorbed in $H(\underline S)!)$.

As to the class of admissible observables we choose them to be real,
differentiable functions with respect to the variables $S(x_i)$ and,
if necessary, bounded away from the internal and (or) external system
boundaries in order to avoid artificial boundary terms in the various
partial integrations. This is however only a measure of precaution
since, typically, at the internal system boundaries they are strongly
supressed in most cases by the exponential vanishing of $exp (-\beta
H)$ for some $S(x)$ approaching $\pm \infty$.

Now, as symmetries induced by Poisson brackets are locally generated
by certain first order partial differential operators, we generalize
the Poisson bracket structure of classical point mechanics by
replacing Poisson brackets with general first order differential
operators acting on the spins in the following way:

\noindent{\bf 2.2 Definition:} A first order differential operator
operating on the spin system is given by
\begin{equation}
{\cal D} := \sum_i d_i (S_i) \partial_{S_{i}}
\end{equation}

with $d_i$ certain twice differentiable functions of the spin
variable $S_i$ ($S_i$ abbreviation for $S(x_i)), \partial _{S_{i}}$
denoting partial differentiation with respect to $S_i$. $\cal D$ is
acting on observables, introduced above, i.e. on certain
differentiable functions of the spins $\{S_i\}$.

\noindent{\bf 2.3 Remark:} Note that the $d_i$'s can in principle
carry an
extra dependence on the site $i$, where they are localized.

With the help of such ${\cal D}$'s we are able to generalize the
KMS-condition (1.1).

\noindent{\bf 2.4 Generalized KMS-Condition:}
\begin{eqnarray}
< {\cal D} A>
& = &<A \cdot (\beta {\cal D} H - \mbox{ div } \underline
d)>\nonumber\\
\mbox{with  div }\underline d
& := & \sum_i \partial_{S_{i}} d_i (S_i)
\end{eqnarray}

and after some calculations the Bogoliubov inequality (1.2).

\noindent{\bf 2.5 Generalized Bogoliubov Inequality:}
\begin{equation}
<{\cal D} A>^2 \; \le \; <A>^2 \cdot <\beta \cdot {\cal D D} H -
{\cal D} (\mbox{ div
}\underline d)>
\end{equation}

\noindent{\bf 2.6 Remarks:} i) That these are the proper extensions
of the
relations (1.1), (1.2) to lattice systems has been shown in [11].

ii) We have given another, slightly different extension in [11] via
augmenting the local phase space at each site $x_i$ and the
Hamiltonian $H$ by means of which we get a true Poisson bracket
structure also on the lattice.

Similar formulas can be derived for spin systems defined over a
continuous base space, i.e.
Euclidean quantum field theory over ${\bf R}^d$, where expectation
values
like (2.1) are replaced by functional integrals and the derivative
operator $\cal D$ by a functional derivative:
\begin{equation}
<A> := Z^{-1} \cdot \int D [\phi] A[\phi] e^{-\beta H [\phi]}
\end{equation}
$D [\phi]$ being the functional measure, $H [\phi]$ the Euclidean
action
\begin{equation}
{\cal D} := \int d^d x \; d(x,\phi (x)) \cdot \delta/\delta \phi (x)
\end{equation}
(cf. [11]).

As one has to cope in this continuum situation with various
renormalisation problems if one does not treat the expressions in a
purely formal manner we plan to study this situation in a more
systematic way in forthcoming work. On the other hand one can try to
carry over the corresponding expressions from the lattice situation
by taking the lattice spacing to zero.

\noindent{\bf 2.7 Remark:} In Euclidean field theory a certain method
has been
in use which bears a weak resemblance to our approach and which is
called ``The Integration by Parts Method''. (cf. e.g. [12]). Our
method however appears to be considerably more general and yields
stronger results since it draws on concepts which have not been
exploited up to now in that field.

\section*{3. Correlation Inequalities for the ${\bf \phi^4}$-System}
\setcounter{equation}{0}
\setcounter{section}{3}

The typical regime of application of relations of the above type is
the important field of phase transitions and spontaneous symmetry
breaking (examples can be found in the above mentioned literature).

In that situation the derivative operator $\cal D$ is typically
chosen to be the generator of the flow representing the continuous
``formal'' symmetry of the model. The attribute ``formal'' means that
in case the symmetry is spontaneously broken the Hamiltonian $H$ is
only invariant under the symmetry in a restricted (formal) sense and
the manipulation of various limiting procedures becomes a highly
delicate matter.

In the following we want to show that, perhaps a little bit
surprisingly, the above formulas can be applied also in a much more
general environment, e.g. where no continuous symmetry exists at all
or where the symmetry is unbroken.

In this wider context the differential operator $\cal D$ does {\bf
not}
suggest itself but has to be chosen cleverly in order to yield
interesting (in)equalities between various expectation values or
correlation functions.

\noindent{\bf 3.1 Remark:} In the following we will choose the
differential
operator $\cal D$ in such a way that $\mbox{div }{\cal D} \equiv 0$
holds in formulas (2.3), (2.4).

To begin with we take as model Hamiltonian the $\phi^4$-Hamiltonian
$(J_{ii} \equiv 0):$
\begin{equation}
H = 1/2 \sum_{ij} J_{ij} S_i S_j  + \sum_i m\; S^2_i + \lambda \sum_i
S^4_i
\end{equation}
and discuss this model for various choices of the parameters.

As interesting and more generic phenomena arise in the thermodynamic
limit $\Lambda \longrightarrow {\bf Z}^d$ the following estimates are
understood in this limit, i.e. they are calculated for finite
$\Lambda$ and are then generalized to the infinite system by a
standard procedure.

The different classes of estimates arise from different choices of the
observable $A$ and operator $\cal D$ in formulas (2.3), (2.4). The
limiting equilibrium state is assumed to represent a ``pure phase''.
This can be achieved in the usual way by adding a symmetry breaking
term in $H$ which is switched off in the end or by fixing certain
boundary conditions. As both $A$ and $\cal D$ are chosen to be local,
i.e. among other things, supported away from the system boundaries,
there remains no explicit effect of these boundary conditions ``at
infinity''.

\noindent{\bf{3.2 The Case $A := S_i, {\cal D} := \partial_{S_{i}}$}}

Inequality (2.4) yields the estimate
\begin{equation}
1 =\; <1>\; \leq \beta <S^2_i> \cdot (2m + 12 \lambda \cdot <S^2_i>)
\end{equation}
with the Hamiltonian (3.1).

Solving for $<S^2_i>$ we get:

i) $m \ge 0:$
\begin{equation}
<S_i^2>\; \ge \;\frac{{\sqrt {\frac{12\lambda}{\beta} + m^2}} - m} {12
\lambda}
\end{equation}

ii) $m = 0:$
\begin{equation}
<S_i^2>\; \ge\; {\sqrt \frac{1}{\beta \cdot 12 \lambda}}
\end{equation}

iii) $m > 0, \lambda \longrightarrow 0:$
\begin{equation}
<S_i^2>\; \ge\; \frac{1}{2m\beta}
\end{equation}

iv) $m < 0 $ (i.e. ground state degenerated):

\noindent From (3.2) we infer:
\begin{equation}
\beta^{-1} \le 12 \lambda \cdot (<S_i^2> + \frac{m}{12 \lambda})^2
-\frac{m^2}{12\lambda}
\end{equation}
and
\begin{equation}
2m + 12 \lambda <S_i^2>\; > 0\qquad \mbox{always}
\end{equation}
which yields
\begin{equation}
<S_i^2> \; > - \frac{m}{6 \lambda} \hspace{1.9cm} \mbox{always}
\end{equation}

Inserting this in (3.6) we see that the bracket is always positive
also for $m<0$!, i.e. there is no problem with the squareroot and we
get, as in the case $m > 0$:
\begin{equation}
<S_i^2>\; \ge\; \frac{{\sqrt {\frac{12\lambda}{\beta} + m^2}} - m} {12
\lambda}, \qquad m < 0
\end{equation}

\noindent{\bf 3.3 Remark:} Note that our estimates hold both for the
case $<S_i> = 0$ and $<S_i>\neq 0$, i.e. with or without spontaneous
magnetization. That is, $<S^2_i>$ is either a pure fluctuation term
or contains the overall magnetization $<S_i>$.

As an application of our above estimates we will study the case $m <
0$ more closely. This is the regime where spontaneous symmetry
breaking becomes possible. It is then an important question for which
values of the phase space parameters, e.g. $\{\beta, m, \lambda\}$ the
equilibrium state of the system represents the non-degenerated phase
(i.e. single phase region) and for what values the equilibrium state
is degenerated (i.e. two pure phases).

\noindent{\bf 3.4 Remark:} At the moment the relation of our results
(presented below) to other kinds of estimates of these bounds (see
e.g. [12]) is not entirely clear to us since they are usually derived
by completely different methods. This point shall be clarified in the
future.

\noindent We now study the particular model Hamiltonian
\begin{eqnarray}
H & = & \frac{1}{2} \sum_{nn} J \cdot (S_i - S_k)^2 + \sum_i (m_o
S^2_i + \lambda S_i^4), \quad J > 0 \nonumber\\
& = & - \sum_{nn} J \cdot S_iS_k + \sum_i (m S^2_i + \lambda S_i^4)
\end{eqnarray}
with $m := m_o + n \cdot J, n =$ number of nearest neighbors.

For $m_0 < 0, J \ge 0$, i.e. ferromagnetic coupling, the ground
state $(\beta = \infty)$, has
$\overline S = \pm \sqrt {\mid m_0 \mid/2\lambda}$. With the help of
our estimate for $<S_i^2>$ we can provide a lower bound of the
strength of fluctuation of $S_i$ and can set it into relation to the
inverse temperature $\beta$. That is:
\begin{eqnarray}
<S^2_i> - \overline{S}^2 & \ge & \overline{S}^2\hspace{2.3cm}
\mbox{implies}\nonumber\\
<(S_i - <S_i>)^2> & \ge & \overline{S}^2  = \frac{\mid m_0 \mid}{2
\lambda}\qquad \mbox{for } 0 \le \;\mid<S_i>\mid \;\le
\mid\overline{S}\mid.
\end{eqnarray}

We can now estimate the critical value for $\beta$ so that the mean
deviation from the average $<S_i>$ becomes larger than
$\mid<S_i>\mid$ or
$\mid\overline{S}\mid$ itself. We conjecture that this signals the
transition
from the two phase to the one phase regime as typical fluctuations
will then connect the two minima. (We, however, do not intend to prove
this at this place).

{}From (3.9) we get:

\noindent{\bf 3.5 Observation:} (3.11) holds if
\begin{equation}
\beta \le \frac{\lambda}{14m^2_0 - 2n J\cdot \mid m_0\mid}=: \beta^*
\qquad (\mbox{ and if } \beta^* > 0!)
\end{equation}
i.e. we suppose that for $\beta \le \beta^*$ the system is in the
one-phase regime.

\noindent{\bf 3.6 The Case $A := S_i \cdot S_j, {\cal D} =
\partial_{S_{j}}$}

Inserting these expressions into (2.4) yields:
\begin{equation}
<S_i>^2 \le \beta <S_i^2 \cdot S_j^2> (2m + 12\lambda <S_j^2>)
\end{equation}

If the system is in a pure state one can exploit well-known cluster
theorems (cf. e.g. [13]) to infer:
\begin{equation}
<S_i^2 \cdot S^2_j> \longrightarrow <S^2_i>\cdot <S_j^2>\; =\;
<S_i^2>^2
\quad \mbox{for } \mid i-j\mid \rightarrow \infty
\end{equation}

This yields:
\begin{equation}
<S_i>^2 \le \beta (2m <S_i^2>^2 + 12 \lambda <S_i^2>^3)
\end{equation}
and for $m < 0 $:
\begin{equation}
<S_i^2>\; \ge\; \left(\frac{<S_i>^2}{\beta \cdot 12
\lambda}\right)^{1/3}
\end{equation}

\noindent{\bf 3.7 The Case $A := S_i^2 \cdot S_j, {\cal D} =
\partial_{S_{j}}$}
\begin{equation}
<S_i^2>^2 \le \beta \cdot <S_i^4 \cdot S_j^2> \cdot (2m + 12 \lambda
<S^2_j>)
\end{equation}
and with the cluster property:
\begin{equation}
<S_i^4>\; \ge\; \frac{<S_i^2>}{\beta (2m + 12 \lambda <S_i^2>)}
\end{equation}

It is evident that one can derive, proceeding in the indicated
manner, a whole sequence of inequalities between various expectation
values and correlation functions.

\noindent{\bf 3.8 Remarks:} i) Up to now our estimates are in general
independent of the strength of the (``kinetic'') coupling $J_{ij}$.
This is a consequence of the choice $\partial_{S_{i}}$ for ${\cal
D}$. For ${\cal D} = \partial_{S_{i}} + \partial_{S_{j}}, i \neq
j$. ${\cal D D} H$ yields also terms containing the couplings
$J_{ij}$.

ii) Furthermore, our results are dimension independent. This is,
however, not always a disadvantage. The dependence on the space
dimension has, on the other side, been exploited in previous work of
us (cited above). To incorporate dimension one has to choose
observables $A$ which go with the volume $\Lambda$. In this paper $A$
was fixed independent of $\Lambda$.

\noindent{\bf 3.9 A Certain Strategy:}

Our estimates yield bounds from below, i.e., as a case in point:
\begin{equation}
<S^2>\; \ge \mbox{ expression (1) in } (\beta, m, \lambda)
\end{equation}

If it is possible to derive a bound of the sort:
\begin{equation}
<(S - <S>)^2> \;\;< \mbox{ expression (2) in } (\beta, m, \lambda)
\end{equation}
it may become possible to get interesting bounds of the kind:
\begin{equation}
<S>\; \neq\; 0\qquad \mbox{for a certain regime of }(\beta, m,
\lambda)
\end{equation}
i.e. estimates concerning the existence of phase transitions.

\noindent{\bf 3.10 Observation:} Note that estimate (3.8) is even
independent of the inverse temperature $\beta$, i.e.
\begin{equation}
<S_i^2>\; > \frac{\mid m\mid}{6 \lambda}\qquad \mbox{for } m < 0
\end{equation}

This implies that in the limit $\beta \rightarrow \infty$ we end up
in a completely ordered phase, i.e. spins aligned, or, in the cases
where an ordered phase is excluded by e.g. Mermin-Wagner-theorem, the
ground state consits of a random occupation of the two minima of the
Hamiltonian.

In closing this paper we would like to point to the fact that up to
now we have only exploited the inequality (2.4). Inserting, on the
other side, the various choices for $A, {\cal D}$ into (2.3) we get
an equality between certain expectation values. For the simplest
choice, i.e.:

\noindent{\bf 3.11 The Case $A := S_i, {\cal D} := \partial_{S_{i}}$;
Equation (2.3)}
\begin{equation}
\beta^{-1} = < \sum_j J_{ij} S_i S_j + 2m S^2_i + 4 \lambda S^4_i >
\end{equation}
which is reminiscent of sort of a virial theorem for spin systems.
Corresponding equations can be derived for other choices of $A, {\cal
D}$.\\[0.5cm]
{\bf 3.12 Remark}: While we think the above scheme is representing a
new approach to this field it may well be that the results can
possibly also be deduced by employing other methods. In any case, an
advantage of our approach is, in our view, its simplicity and
transparency.
\vspace{1cm}

\noindent {\bf Acknowledgement:} Several fruitful discussions with
H.J. Wagner are gratefully acknowledged.


\begin{thebibliography}{99}
\bibitem{}
M. Requardt: Zeitschr. Phys. {\bf B 36} (1979) 187
\bibitem{}
Ph. A. Martin: Nuovo Cimento {\bf B 68} (1982) 302
\bibitem{}
M. Requardt: Journ. Stat. Phys. {\bf 31} (1983) 679
\bibitem{}
M. Requardt; H.J. Wagner: Journ. Stat. Phys. {\bf 45} (1986)  815
\bibitem{}
H.J. Wagner: Physica {\bf A 144} (1987) 495
\bibitem{}
M. Requardt: Journ. Stat. Phys. {\bf 50} (1988) 737
\bibitem{}
M. Requardt; H.J. Wagner: Physica {\bf A 154} (1988) 183
\bibitem{}
M. Requardt: Zeitschr. Phys. {\bf B 73} (1988) 133
\bibitem{}
G. Gallavotti; E. Verboven: Nuovo Cimento {\bf B 28} (1975) 274
\bibitem{}
N.D. Mermin: Journ. Math. Phys. {\bf 8} (1967) 1061
\bibitem{}
M. Requardt; H.J. Wagner: Phys. Lett. {\bf A} (1989) 303
\bibitem{}
J. Glimm; A. Jaffe: Quantum Physics, A Functional Integral Point of
View, Springer, N.Y. 1981
\bibitem{}
D. Ruelle: Statistical Mechanics, Rigorous Results, Benjamin Inc.,
N.Y. 1969
\end{thebibliography}
\end{document}